\documentstyle[12pt]{article}
\def\beq{\begin{equation}}
\def\eeq{\end{equation}}
\def\bea{\begin{eqnarray}}
\def\eea{\end{eqnarray}}

\def\ba{\begin{array}}
\def\ea{\end{array}}
\def\bce{\begin{center}}
\def\ece{\end{center}}

\def\nonu{\nonumber}

\def\pa{\partial}

\begin{document}
\begin{titlepage}
\rightline{SNUTP-97-171}
\rightline{hep-th/9712149}
\def\today{\ifcase\month\or
January\or February\or March\or April\or May\or June\or
July\or August\or September\or October\or November\or December\fi,
\number\year}
\vskip 1cm
\centerline{\Large \bf Confining Phase of $N=1$ $Sp(N_c)$ Gauge Theories }
\centerline{\Large \bf  via M Theory Fivebrane }
\vskip 2cm
\centerline{\sc Changhyun Ahn$$\footnote{ chahn@spin.snu.ac.kr}} 
\vskip 1cm
\centerline{{\it  Dept. of Physics,}}
\centerline{ {\it Seoul National University, }}
\centerline{ {\it Seoul 151-742, Korea}}
\vskip 4cm
\centerline{\sc Abstract}
\vskip 0.2in
The moduli space of vacua for the confining phase of $N=1$ $Sp(N_c)$ supersymmetric
gauge theories in four dimensions is studied by M theory fivebrane.
We construct M theory fivebrane configuration corresponding to the perturbation of
superpotential in which the power of adjoint field is related to the number of
NS'5 branes in type IIA brane configuration. We interpret the dyon vacuum expectation 
values in field theory results as the brane geometry and the comparison with meson 
vevs will turn out that the low energy effective superpotential with enhanced gauge
group $SU(2)$ is exact.

\vskip 1in
\leftline{Dec., 1997}
\end{titlepage}
\newpage
\setcounter{equation}{0}

\section{Introduction}
\setcounter{equation}{0}

We have seen how string/M theory can be exploited to study 
non-perturbative dynamics of  low energy  supersymmetric gauge theories in
various dimensions. 
One of the main motivations is to understand the D
brane dynamics where the gauge theory is realized on the worldvolume of D branes.
This work was initiated by Hanany and Witten \cite{hw} where 
the mirror symmetry of $N=4$ gauge theory in three dimensions was described
by changing the position of the NS5 branes  ( See,
for example,  \cite{bo1}  ).
As one changes the relative orientation of the two NS5 branes \cite{bar} while keeping
their common four spacetime dimensions intact, the $N=2$ supersymmetry is
broken to $N=1$ supersymmetry \cite{egk}. 
By analyzing this brane configuration they  \cite{egk} described and checked a stringy
derivation of Seiberg's duality for $N=1$ supersymmetric $SU(N_c)$ gauge theory with
$N_f$ flavors.
This result was generalized to the brane configurations with orientifolds which 
explain $N=1$ supersymmetric theories with gauge group $SO(N_c)$ or 
$Sp(N_c)$ \cite{eva} ( See also \cite{ov} for a relevant geometrical approach ).

Both the D4 branes
and NS5 branes used in type IIA string theory 
originate from the fivebrane \cite{w1} of M theory. That is, D4 brane is an M theory
fivebrane wrapped over $\bf{S^1}$ and NS5 brane is the one on
$ \bf{R^{10} \times S^1} $. In order to insert  D6 branes 
one studies a multiple
Taub-NUT space \cite{town}
whose metric is complete and smooth.
The singularities 
are removed in eleven dimensions where the brane configuration becomes smooth, the
D4 branes and NS5 branes being the unique M theory  fivebrane and 
the D6 branes being the Kaluza-Klein monopoles.
The property of $N=2$ supersymmetry 
in four dimensions requires that the worldvolume of M theory fivebrane is 
${\bf{R^{1,3}}} \times \Sigma$ where $\Sigma$ is uniquely identified with the
curves that 
occur to the solutions for Coulomb branch of the four dimensional field theory.
Further generalizations of this configuration with orientifolds were 
studied in \cite{lll}.  
The exact low energy description of $N=1$ supersymmetric $SU(N_c)$ gauge
theories with $N_f$ flavors in four dimensions have been found in \cite{w2} ( See
also \cite{aotaug,aotsept} for theories with orientifolds ).   
This approach has been developed further and
used to study the moduli space of vacua of confining phase of $N=1$ supersymmetric
$SU(N_c)$ gauge theories in four dimensions \cite{dbo}. 
In terms of brane configuration of IIA string
theory, this was done by taking 
multiples of NS'5 branes rather than a single NS'5 brane. In field theory, we
regard this as taking the superpotential
$\Delta W = \sum_{k=2}^{N_{c}} \mu_{k} \mbox{Tr} (\Phi^{k})$. This
perturbation lifts the non singular locus of the $N=2$ Coulomb branch while
at singular locus there exist massless dyons that can condense due to the perturbation.

In the present work we extend the results of \cite{dbo,aotdec} to 
$N=1$ supersymmetric theories
with gauge group $Sp(N_{c})$ and also generalize the previous
work \cite{aotaug} which dealt with a single NS'5 brane  
in the sense that we are considering {\it multiple copies} of NS'5 branes. 
We will describe how the field theory analysis \cite{ty} 
obtained in the low energy superpotential
gives rise to the geometrical structure in $(v, t, w)$ space.
The minimal form for the effective
superpotential obtained by ``integrating in" is not exact \cite{intril},
in general, for several massless dyons.
Note that
the intersecting branes in string/M theory
have been studied to obtain much information about 
supersymmetric gauge theories with different gauge groups and in various 
dimensions \cite{ah}. 

\section{Field Theory Analysis}
\setcounter{equation}{0}

$\bullet$  $N=2$ Theory

Let us consider $N=2$ supersymmetric  $Sp(N_{c})$ gauge theory
with matter in the ${\bf 2 N_c}$ 
dimensional representation of the gauge group $Sp(N_c)$. In terms
of $N=1$ superfields, $N=2$ vector multiplet consists of a field strength
chiral multiplet $W_{\alpha}^{ab}$ and a scalar chiral multiplet 
$\Phi_{ab}$, both in the
adjoint representation. The quark hypermultiplets
are made of a chiral multiplet $Q^{i}_{a}$ which couples to the 
Yang-Mills fields where 
$i = 1,\cdots ,2N_{f}$ are flavor indices
and $a = 1, \cdots , 2N_{c}$ are color indices. 
The
$N=2$ superpotential takes the form,
\begin{equation}
\label{super}
W_{N=2} = \sqrt{2} Q^{i}_{a} \Phi^a_b J^{bc} Q^{i}_{c} 
+ \sqrt{2} m_{ij} Q^{i}_{a} J^{ab} Q^{j}_{b},
\end{equation}
where $J_{ab}$ is the symplectic metric 
$( {0 \atop -1 }{ 1 \atop 0}  ) \otimes {\bf 1_{N_c \times N_c}} $
used to raise and lower 
$Sp(N_{c})$ color indices ( $
{\bf 1_{N_c \times N_c}}$ is the $N_c \times N_c$ identity matrix )
and $m_{ij}$ is an antisymmetric quark mass matrix
$\label{mass}( { 0 \atop 1 }{  -1 \atop 0 }  ) 
\otimes \mbox{diag} ( m_{1}, \cdots, m_{N_f} ) $.
Classically, the global symmetries are the flavor
symmetry $O(2N_{f})$ when there are no quark masses, in addition to  
$U(1)_{R}\times SU(2)_{R}$ chiral R-symmetry. 
The theory is asymptotically free for the region $N_{f} < 2N_{c}+2$ and generates
dynamically a strong coupling scale $\Lambda_{N=2}$.
The instanton factor
is proportional to $\Lambda_{N=2}^{2N_{c}+2-N_{f}}$. Then the 
$U(1)_{R}$ symmetry is anomalous and is broken down to a discrete 
$\bf{Z_{2N_{c}+2-N_{f}}}$ symmetry by instantons.
The $N_c$ complex dimensional moduli space of vacua 
contains the Coulomb and Higgs branches.
The Coulomb branch is parameterized by the gauge invariant order parameters
\bea
u_{2k}=<\mbox{Tr}(\phi^{2k})>, \;\;\;\;\; k=1, \cdots, N_c,
\label{u2k}
\eea 
where $\phi$ is the scalar field in $N=2$ chiral multiplet.
Up to a gauge transformation $\phi$  can be  diagonalized to a
complex matrix, 
$<\phi>=\mbox{diag} ( A_1, \cdots, A_{N_c} )$ where $A_i=
( { a_i \atop  0 }{  0 \atop  -a_i}  )$.
At a generic point the vevs of 
$\phi$ breaks the $Sp(N_c)$ gauge symmetry
to $U(1)^{N_c}$ and the dynamics of the theory is that of an 
Abelian Coulomb phase. The Wilsonian effective Lagrangian in the low
energy can be made of the multiplets of $A_i$ and $W_i$ where
$i=1, 2, \cdots, N_c$. 
If $k$ $a_i$'s are equal and nonzero then there 
exists an enhanced $SU(k)$ gauge symmetry. When they are also zero, 
an enhanced $Sp(k)$ gauge symmetry appears.
The quantum moduli space is described by a family of 
hyperelliptic spectral curves \cite{aps}
with associated 
meromorphic one forms,
\bea 
y^2  = \left( v^2 C_{2N_c}(v^2)+\Lambda_{N=2}^{2N_c+2-N_f} 
\prod_{i=1}^{N_f} m_i \right)^2-
\Lambda_{N=2}^{4N_c+4-2N_f}  \prod_{i=1}^{N_f}(v^2-m_i^2),
\label{curve}
\eea
where $C_{2N_c}(v^2)$ is a degree $2N_c$ polynomial in $v$ with  coefficients
depending on the moduli $u_{2k}$ appearing in (\ref{u2k})  and 
$m_i ( i=1, 2, \cdots, N_f )$ is the
mass of quark\footnote{
Note that the polynomial $C_{2N_c}(v^2)$ is an even function of $v$ which will be identified 
with a complex coordinate $( x^4, x^5 ) $ directions in next section and 
is given by 
$C_{2N_c}(v^2)= v^{2N_c} +\sum_{i=1}^{N_c} s_{2i} v^{2(N_c-i)}=
\prod_{i=1}^{N_c} (v^2-a_i^2) $
where $s_{2k}$ and $u_{2k}$ are related each other  by so-called Newton's formula
$2k s_{2k}+ \sum_{i=1}^{k} s_{2k-2i} u_{2i} =0, ( k=1, 2, \cdots, N_c )$
with $s_0=1$. From this recurrence relation,
we obtain $ \pa s_{2j}/\pa u_{2k}=- s_{2(j-k)}/2k
\;\;\; \mbox{for} \;\;\; j \geq k. $}.

$\bullet$ Breaking $N=2$ to $N=1$ ( Pure Yang-Mills Theory )

We are interested in a microscopic $N=1$ theory mainly in a phase
with a single confined photon
coupled to the light dyon hypermultiplet while the photons for the rest are free.
By taking a tree level superpotential perturbation  
$\Delta W$ of \cite{ty} made out of the adjoint fields in the vector
multiplets to the $N=2$ 
superpotential (\ref{super}), the $N=2$ supersymmetry 
can be broken to $N=1$ supersymmetry.  That is,
\bea
W = W_{N=2}+\Delta W, \;\;\;
\Delta W  =  \sum_{k=1}^{N_c-1} \mu_{2k} \mbox{Tr} (\Phi^{2k}) +
\mu_{2N_c} s_{2N_c},
\label{superpo} 
\eea
where\footnote{ Our $\mu_{2k}$ is the same as their $g_{2k}/2k$ in \cite{ty}. } 
$\Phi$ is the adjoint $N=1$ superfields in the $N=2$ vector multiplet and
Note that the $\mu_{2N_c}$ term is not associated with $u_{2N_c}$ but
$s_{2N_c}$ which is proportional to the sum of $u_{2N_c}$ and the polynomials
of other $u_{2k} ( k < N_c )$ according to the recurrence relation.
Then microscopic $N=1$ $Sp(N_c)$ gauge theory is obtained from $N=2$ $Sp(N_c)$
Yang-Mills theory perturbed by $\Delta W$.
Let us first study $N=1$ pure $Sp(N_c)$ Yang-Mills theory with tree level 
superpotential (\ref{superpo}).
Near the singular points where dyons become massless,
the macroscopic superpotential of the theory is given by
\bea
W = \sqrt{2} \sum_{i=1}^{N_c-1} M_i A_i M_i + \sum_{k=1}^{N_c-1}
\mu_{2k} U_{2k} +\mu_{2N_c} S_{2N_c}.
\label{supereven}
\eea
We denote by $A_i$ the $N=2$ chiral superfield of $U(1)$
gauge multiplets, by $M_i$ those of $N=2$ dyon hypermultiplets, 
by $U_{2k}$ the chiral superfields corresponding to $\mbox{Tr} (\Phi^{2k}) $ ( and 
by $S_{2k}$ the chiral superfields which are related to $U_{2k}$ ), 
in the low energy theory. The vevs of the lowest
components of $A_i, M_i, U_{2k}, S_{2N_c} $ are written as 
$a_i, m_{i, dy}, u_{2k}, s_{2N_c}$
respectively.
Recall that $N=2$ configuration is invariant under the group $U(1)_R$
and $SU(2)_R$ corresponding to the chiral R-symmetry of the field theory. 
However,
in $N=1$ theory $SU(2)_R$ is broken to $U(1)_J$. 
The equations of motion obtained by varying the superpotential with respect to 
each field read as follows\footnote{ $
\mu_{2k}=  
-\sqrt{2} \sum_{i=1}^{N_c-1}  m_{i,dy}^2 \pa a_i/\pa u_{2k},
\mu_{2N_c}  =  
-\sqrt{2} \sum_{i=1}^{N_c-1}  m_{i,dy}^2 \pa a_i/\pa s_{2N_c} $
and $ a_i m_{i, dy}=0.$}.
At a generic point in the moduli space, no massless fields occur (
$a_i\neq 0$ for $ i=1, \cdots, N_c-1 $ ) which
implies $m_{i, dy} =0$ and so $\mu_{2k}$ and $ \mu_{2N_c}$
vanish.  Then we obtain
the moduli space of vacua of $N=2$ theory. 

On the other hand, we are considering a singular point in the moduli space where
$l$ mutually local dyons are massless. 
This means that $l$ one cycles
shrink to zero. 
The right hand side of (\ref{curve}) becomes,    
\bea
y^2 = 
\left( v^2 C_{2N_c}(v^2)+\Lambda_{N=2}^{2N_c+2-N_f}  \right)^2-
\Lambda_{N=2}^{4N_c+4-2N_f}  =
\prod_{i=1}^{l}(v^2-p_i^2)^2
\prod_{j=1}^{2N_c+2 -2l}(v^2-q_j^2),
\label{curve1}
\eea
with $p_i$ and $q_j$ distinct. 
A point in the $N=2$ moduli space of vacua is characterized by $p_i$ and $q_j$.
The degeneracy of this curve is checked by explicitly
evaluating both $y^2$ and $\pa y^2/ \pa v^2$ 
at the point $v= \pm p_i,$ leading to vanish.
Since $a_i =0$ for $i=1, \cdots, l$  and $a_i\neq 0$ for 
$ i=l+1, \cdots, N_c-1 $,
we get $
m_{i, dy}=0, ( i = l+1, \cdots , N_c-1 ) $
while $m_{i, dy} ( i=1, \cdots, l )$ are not constrained.
We will see how the vevs  $m_{i, dy}$ originate from the information about
$N=2$ moduli space of vacua which is encoded by $p_i$ and $q_j$.
We assume that the matrix  $\pa a_i/ \pa u_{2k}$ is nondegenerate.
In order to calculate $\pa a_i/ \pa u_{2k}$,
we need the relation
between $ \pa a_i/ \pa s_{2k}$ and the period integral
on a basis of holomorphic one forms on the curve, $
\pa a_i/\pa s_{2k}= \oint_{\alpha_i} v^{2(N_c-k)}dv/y $.
We can express the generating function\footnote{
By plugging $y$ of (\ref{curve1})  and  integrating along one cycles
around $v=\pm p_i $, it turns out $
\pa a_i/\pa s_{2k} = p_i^{2(N_c-k)}/
\prod_{j \neq i}^l (p_i^2-p_j^2)\prod_t^{2N_c+2-2l}(p_i^2-q_t^2)^{1/2}, $
since the $l$ one cycles shrink to zero.
Then,
we arrive at the following relation,
$ 2k \mu_{2k} = 
\sum_{i=1}^l \sum_{j=1}^{N_c} s_{2(j-k)} p_i^{2(N_c-j)}
\omega_i,$ where $\omega_i = \sqrt{2} m_{i,dy}^2/
\prod_{s \neq i}^l (p_i^2-p_s^2) \prod_t^{2N_c+2-2l} (p_i^2-q_t^2)^{1/2}.$}
for the $\mu_{2k}$ in terms of $\omega_i$
as follows:
\bea
\sum_{k=1}^{N_c} 2k \mu_{2k} v^{2(k-1)} & = & \sum_{k=1}^{N_c} \sum_{i=1}^l 
\sum_{j=1}^{N_c} v^{2(k-1)} s_{2(j-k)} p_i^{2(N_c-j)} \omega_i \nonu \\
& = & 
\sum_{k=-\infty}^{N_c} 
\sum_{i=1}^l 
\sum_{j=1}^{N_c} v^{2(k-1)} s_{2(j-k)} p_i^{2(N_c-j)} \omega_i  + 
{\cal O}(v^{-2}) \nonu \\
& = &  \sum_{i=1}^l  \frac{ C_{2N_c}(v^2)}{(v^2-p_i^2)} \omega_i 
+ {\cal O}(v^{-2}).
\label{mu2k}
\eea
Therefore we can find the parameter $\mu_{2k}$ by reading off the right hand side
of (\ref{mu2k}).
This result determines whether a point in the $N=2$ moduli space of vacua
classified by the set of $p_i$ and $ q_j$ in (\ref{curve1}) remains as an $N=1$ vacuum
after the perturbation, for given a set of perturbation parameters 
$\mu_{2k}$ and $ \mu_{2N_c}$.
We will see in section 3 that this corresponds to one of the boundary conditions on
a complex coordinate in $( x^8, x^9 )$ directions as 
$v$  goes to infinity.
In order to make the comparison with the brane picture, 
it is very useful to define the polynomial
$H (v^2)$ of degree $2l-4$ by
\bea
\sum_{i=1}^l \frac{\omega_i}{v^2(v^2-p_i^2)} = 
\frac{2H (v^2)}{ \prod_{i=1}^l (v^2-p_i^2)}.
\label{heven}
\eea
At a given point $p_i$ and  $q_j$ in the $N=2$ moduli space of vacua, 
$H (v^2)$ determines the dyon vevs, in other words,
\bea
m_{i,dy}^2 = \sqrt{2} p_i^2 H (p_i^2) \prod_m (p_i^2-q_m^2)^{1/2},
\label{dyoneven}
\eea
which will be described in terms of the geometric brane picture in next section.
Therefore, all the vevs of dyons $m_{i, dy} ( i=1, \cdots, l )$ are found as well as
$m_{i, dy} ( i=l+1, \cdots, N_c-1 )$ which are zero.

$\bullet$ The Meson Vevs 

Let us discuss the vevs of the meson field along the singular locus of the Coulomb
branch. This is due to the nonperturbative effects of $N=1$ theory and obviously
was zero before the perturbation (\ref{superpo}). We will see the property of exactness
in field theory analysis in the context of M theory fivebrane in section 4. Equivalently,
the exactness of superpotential for any values of the parameters is to assume 
$W_{\Delta}=0$.
We will follow the method presented in \cite{efgir}.
Let us consider the vacuum where one massless dyon exists 
with unbroken $SU(2) \times U(1)^{N_c-1}$ where
\bea
J \Phi^{cl}= \left( { 0 \atop 1 }{  1 \atop 0 }  \right) \otimes 
\mbox{diag} (a_1, a_1, a_2, \cdots, a_{N_c-1}),
\label{vacuaeven}
\eea
and the chiral multiplet $Q=0$ and $J$ is the symplectic metric as before.
These eigenvalues of $\Phi$ can be obtained by differentiating the superpotential
(\ref{superpo}) with respect to $\Phi$ and setting the chiral multiplet $Q=0$.
The vacua with classical $SU(2) \times U(1)^{N_c-1}$ group are those with
two eigenvalues equal to $a_1$ and the rest given by $a_2, a_3, \cdots, a_{N_c-1}$.
It is known from \cite{ty} that, if using $s_{2N_c}$ in the superpotential
perturbation rather than $u_{2N_c},$
the degenerate eigenvalue of $\Phi$
is obtained to be
\bea
a_1^2=\frac{(N_c-1) \mu_{2(N_c-1)}}{N_c \mu_{2N_c}}.
\label{a1}
\eea
We will see in section 3 that 
the asymptotic behavior of a complex coordinate in $( x^8, x^9 )$ directions for large $v$
determines this degenerate eigenvalue by using the condition for generating function of
$\mu_{2k}$ (\ref{mu2k}).
The scale matching condition between the high energy $Sp(N_c)$ scale
$\Lambda_{N=2}$ and the low energy $SU(2)$ scale $\Lambda_{SU(2), N_{f}}$
is related each other.
After integrating out $SU(2)$ quarks we obtain the scale matching
between $\Lambda_{N=2}$ and $\Lambda_{SU(2)}$ for pure $N=1$ $SU(2)$
gauge theory. That is,
\bea
\Lambda_{SU(2)}^6=\left( \frac{2N_c^2 \mu_{2N_c}^2}{(N_c-1) \mu_{2(N_c-1)}} \right)^2 
\Lambda_{N=2}^
{4(N_c+1)-2N_f} \mbox{det} (a_1^2-m^2),
\eea 
where matrix $a_1$ means
$( { 0 \atop 1 }{  1 \atop 0 }  )  \otimes a_1$ and  
matrix $m$ being 
$( { 0 \atop 1 }{  -1 \atop 0 }  ) 
\otimes \mbox{diag} ( m_{1}, \cdots, m_{N_f} ) $.
Then the full exact low energy effective superpotential is given by
\bea
W_L = 
\sum_{k=1}^{N_c-1} \mu_{2k} \mbox{Tr} (\Phi_{cl}^{2k}) +
\mu_{2N_c} s^{cl}_{2N_c}
\pm 2 \Lambda_{SU(2)}^3,
\label{lowsuperpo}
\eea
where the last term is generated by
gaugino condensation in the low energy $SU(2)$ theory ( the sign reflects the vacuum
degeneracy ).
In terms of the original $N=2$ scale, we can write it as
\bea
W  =  \sum_{k=1}^{N_c-1} \mu_{2k} \mbox{Tr} (\Phi_{cl}^{2k})  
+\mu_{2N_c} s^{cl}_{2N_c} \pm \frac{4N_c^2 \mu_{2N_c}^2}{(N_c-1) \mu_{2(N_c-1)}}
\Lambda_{N=2}^{2(N_c+1)-N_f} \mbox{det} (a_1^2-m^2)^{1/2}.
\eea
Therefore,
one obtains the vevs of meson
$M_i=Q^i_a  J^{ab} Q^i_b$  which gives
\bea
M_i=\frac{\pa W}{\pa m_i^2}=\pm \frac{4 \Lambda_{N=2}^
{2(N_c+1)-N_f}}{\sqrt{2}(a_1^2-m_i^2)}
\frac{N_c^2 \mu_{2N_c}^2}{(N_c-1) \mu_{2(N_c-1)}}  \mbox{det} (a_1^2-m^2)^{1/2},
\label{meson}
\eea
where $a_1$ is given by ({\ref{a1}).
We will see in section 3 that the finite value of a complex coordinate 
in $( x^8, x^9 )$ directions corresponds to the above vevs of meson
when $v \rightarrow \pm m_i$ and the other complex coordinate related to $( x^6, x^{10} )$ 
directions vanishes.

\section{ Brane Configuration from M Theory }
\setcounter{equation}{0}

$\bullet$ Type IIA Brane Configuration 

We study the theory with the superpotential perturbation
$\Delta W$ (\ref{superpo}) by analyzing M theory fivebranes. 
Let us first describe them in the type IIA brane configuration.
Following the procedure of \cite{egk}, the brane
configuration in $N=2$ theory consists of
three kind of branes: the two parallel NS5
branes extend in the directions $(x^0, x^1, x^2, x^3, x^4, x^5)$, the D4 branes
are stretched between two NS5 branes and extend over $(x^0, x^1, x^2, x^3)$ and
are finite in the direction of $x^6$, and the D6 branes extend in the directions
$(x^0, x^1, x^2, x^3, x^7, x^8, x^9)$. 
In order to study 
symplectic gauge groups, we  consider an O4
orientifold which is parallel to the D4 branes in order to keep the
supersymmetry and is not of finite extent in $x^6$ direction. The D4 branes
is the only brane which is not intersected by this O4 orientifold. 
The orientifold
gives a spacetime reflection as $(x^4, x^5, x^7, x^8, x^9) \rightarrow
(-x^4, -x^5, -x^7, -x^8, -x^9)$, in addition to the gauging of worldsheet
parity $\Omega$. 
The fixed points of the spacetime symmetry define this O4 planes.
Each object which does not lie at the fixed points, 
must have its mirror. Thus NS5 branes have a mirror in $(x^4,
x^5)$ directions and D6 branes do a mirror in $(x^7, x^8, x^9)$ directions.
In order to realize the $N=1$ theory with a perturbation (\ref{superpo})
we can think of a single NS5 brane and {\it multiple copies } of NS'5 branes
which are orthogonal to a NS5 brane 
with worldvolume, $(x^0, x^1, x^2, x^3, x^8, x^9)$ and between them there
exist D4 branes intersecting D6 branes. The number of NS'5 branes is
$N_c-1$ by identifying the power
of adjoint field appearing in the superpotential (\ref{superpo}).
The brane description for $N=1$ theory with $\mu_{2N_c}=0$
has been studied in the paper of \cite{egk} in type IIA brane configuration. In this case,
all the couplings, $\mu_{2k}$ can be regarded as tending uniformly to infinity. On the other
hand, in M theory configuration there will be no such restrictions.

$\bullet$   M Theory Fivebrane Configuration 

Let us describe how the above brane configuration can be embedded in 
M theory in terms of
a single M theory fivebrane whose worldvolume is
${\bf R^{1,3}} \times \Sigma$ where $\Sigma$ is identified with Seiberg-Witten
curves \cite{aps} that determine  the solutions to Coulomb branch 
of the field theory.
As usual, we write $s=(x^6+i x^{10})/R, t=e^{-s}$
where $x^{10}$ is the eleventh coordinate of M theory which is compactified
on a circle of radius $R$. Then the curve $\Sigma$, describing
$N=2$  $Sp(N_c)$ gauge theory with $N_f$ flavors,
is given \cite{lll} by an equation in $(v, t)$ space
\bea
t^2 - 2 \left( v^2 C_{2N_c}(v^2, u_{2k})+\Lambda_{N=2}^{2N_c+2-N_f} \prod_{i=1}^{
N_f} m_i \right) t  + 
\Lambda_{N=2}^{4N_c+4-2N_f}
\prod_{i=1}^{N_f} (v^2 -m_i^2) = 0.
\label{cur}
\eea 
It is easy to check that this description is the same as (\ref{curve})
under the identification $
t= y + v^2 C_{2N_c}(v^2, u_{2k})+\Lambda_{N=2}^{2N_c+2-N_f} \prod_{i=1}^{N_f} m_i. $
By adding $\Delta W$ which corresponds to the adjoint chiral multiplet, 
the $N=2$ supersymmetry  will be broken to $N=1$. 
To describe the corresponding brane configuration in M theory,
we introduce  complex coordinates
\bea
v= x^4 +i x^5, \;\;\; w=x^8+i x^9.
\eea
To match the superpotential perturbation $\Delta W $ (\ref{superpo}),
we propose the following boundary conditions 
\bea
\begin{array}{ccl}
w^{2} & \rightarrow & \sum_{k=2}^{N_c} 2k \mu_{2k} v^{2(k-1)}\;\;\;\mbox{as}~
v \rightarrow \infty,~
t \sim \Lambda_{N=2}^{2(2N_c+2-N_f)}v^{2N_f-2N_c-2},\\
w & \rightarrow & 0\;\;\;\mbox{as}~
v \rightarrow \infty,~
t \sim v^{2N_c+2}. \\
\end{array}
\label{boundary}
\eea
After deformation, $SU(2)_{7,8,9}$ is broken to $U(1)_{8,9}$ if $\mu_{2k}$
has the charges $(4-4k, 4)$ under $U(1)_{4,5} \times U(1)_{8,9}$.
When we consider now only $k=2$, we obtain that 
$w^2 \sim \mu_{4} v^{2}$ as $v \rightarrow \infty$ which is
the same as the relation $w \rightarrow \mu v$  in 
\cite{aotaug} after we identify $\mu_{4}$ with $\mu^{2}$. This 
identification comes also from the consideration of $U(1)_{4,5}$ and $U(1)_{8,9}$ 
charges of $\mu$ and $\mu_{4}$. 
 After perturbation, only the singular part of the $N=2$ Coulomb branch with
$l$ or less mutually local massless dyons remains in the moduli space of vacua.
Let us construct the M theory fivebrane configuration satisfying
the above boundary conditions and assume 
that $w^{2}$ is a rational function of $v^{2}$ and $t$.
Our result is really similar to the case of \cite{dbo,aotdec} and we
will follow their notations.
We  write $w^{2}$ as follows
\bea
w^{2}(t,v^2) = \frac{a(v^2) t + b(v^2)}{c(v^2) t + d(v^2)},
\eea
where $a, b, c, d$ are arbitrary polynomials of $v^2$ and $t$ satisfying  
(\ref{cur}). 
Now we calculate the following two quantities\footnote{ Using the two
solutions of $t$, denoted by $t_+$ and $t_-$ satisfying (\ref{cur}), $
w^{2}(t_+(v^2),v^2)+w^{2}(t_-(v^2),v^2) = (2acG+2adC+2bcC+2bd)/( c^2 G + 2 cdC + d^2)
$
and $
w^{2}(t_+(v^2),v^2)-w^{2}(t_-(v^2),v^2) = 2(ad-bc)S\sqrt{T}/(c^2 G +2 cdC + d^2)
$
where  we define $
C \equiv v^2 C_{2N_c}(v^2, u_{2k})+\Lambda_{N=2}^{2N_c+2-N_f} 
\prod_{i=1}^{N_f} m_i, $
and $ G \equiv \Lambda_{N=2}^{4N_c-4-2N_f}
\prod_{i=1}^{N_f} (v^2 -m_i^2) $
implying that $
C^2(v^2) -  G(v^2) \equiv S^2(v^2) T(v^2)  $
where $
S(v^2)=\prod_{i=1}^{l} (v^2-p_i^2), T(v^2)=\prod_{j=1}^{2N_c+2-2l} (v^2-q_j^2) $
with all $p_i $ and  $ q_j$'s different.}.
Since $w^{2}$ has no poles for finite value of $v^2$,
$w^{2}(t_+(v^2),v^2) \pm w^{2}(t_-(v^2),v^2)$ also does not have poles which leads to
arbitrary polynomials $H(v^2)$ and $N(v^2)$ given by
\bea 
\frac{acG +  adC + bcC +bd}{c^2 G + 2  cdC + d^2}  =  N,  \;\;\;
\frac{(ad-bc)S}{(c^2 G +2 cdC + d^2)}  =  H.
\eea
It will turn out that the function $H(v^2)$ is exactly 
the same as the one (\ref{heven}) 
defined in field theory analysis.
By making a shift of $a \rightarrow a +Nc, b \rightarrow b +Nd$ due to the 
arbitrariness of the polynomials $a$ and $b$  and
combining all the information for $b$ and $d$ ( See, for details, \cite{dbo,aotdec} ), we get
the most general rational function $w^2$ which has no poles for finite
value of $v^2$,
\bea 
w^{2} = N + \frac{at + cHST - aC}{ct - c C + aS/H},
\eea
where $N, a, c, H$ are arbitrary polynomials.
As we choose two $w^2$'s , each of them possessing different polynomials $a$ and $c$
and subtract them, then the numerator of it will be proportional to $t^2-2 C t+G$ which
vanishes according to (\ref{cur}). This means $w^2$ does not depend on $a$ and $c$.
Therefore,
when $c=0$, the form of $w^2$
is very simple. 
The general solution for $w^2$ is 
\bea 
w^{2} = N(v^2) +  \frac{H(v^2) }{\prod_{i=1}^{l}(v^2-p_i^2)} 
 \left( t- \left( v^2 C_{2N_c}(v^2, u_{2k})+\Lambda_{N=2}^{2N_c+2-N_f}
\prod_{i=1}^{N_f} m_i \right)  \right),
\eea
where $H(v^2)$ and $N(v^2)$ are arbitrary polynomials of $v^2$.
Now we want to impose the boundary conditions on $w^2$ in
the above general solution.  From the relation, 
\bea 
w^{2}(t_{\pm}(v^2),v^2) = N \pm H\sqrt{T},
\label{nht}
\eea
when we impose the boundary condition 
$ w \rightarrow 0 $ for  $v \rightarrow \infty,~
t = t_-(v) \sim v^{2N_c+2}$
the polynomial $N(v^{2})$ is determined as follows,
\bea
N(v^2)=\left[  H(v^2) \sqrt{T(v^2)} \right]_+=
\left[  H(v^2) \prod_{j=1}^{2N_c+2-2l}(v^2-q_j^2)^{1/2} \right]_+,
\eea 
where $\left[ H(v^2) \sqrt{T(v^2)} \right]_+$ means only nonnegative power of $v^2$
when we expand around $v = \infty$.
The other boundary condition tells that $w^2$ behaves
as
$ w^{2} \rightarrow \sum_{k=1}^{N_c} 2k \mu_{2k} v^{2(k-1)} $ from (\ref{boundary}).
Then by expanding $w^2$ in powers of $v^2$ we can identify $H(v^2)$ with
parameter $\mu_{2k}$. Using 
$T^{1/2}= (t-( v^2 C_{2N_c}(v^2, u_{2k})+\Lambda_{N=2}^{2N_c+2-N_f} \prod_{i=1}^{
N_f} m_i  ) )/S$ and $t=2 ( v^2 C_{2N_c}+ \cdots )$ from (\ref{cur}) we get
\bea
w^{2} & =& \left[2  H(v^2) \sqrt{T(v^2)} \right]_+ + {\cal O}(v^{-2}) \nonu \\
& = & \frac{2 H(v^2)}{\prod_{i=1}^{l}(v^2-p_i^2)}
\left( v^2 C_{2N_c}(v^2, u_{2k})+\Lambda_{N=2}^{2N_c+2-N_f} 
\prod_{i=1}^{N_f} m_i  \right)
+ {\cal O}(v^{-2})  \nonu \\
& = &  
\sum_{i=1}^l \frac{ C_{2N_c} (v^2) \omega_i}{(v^2-p_i^2)} + {\cal O}(v^{-2}) =
\sum_{k=1}^{N_c} 2k \mu_{2k} v^{2(k-1)},
\eea
where we used the definition of $H $ in (\ref{heven})
and the generating function of $\mu_{2k}$ in (\ref{mu2k}). From this result
one can find the explicit form of $H(v^2)$ in terms of $\mu_{2k}$ by comparing both sides
in the above relation. This is an explanation for field theory results 
of (\ref{mu2k}) and (\ref{heven}) which determine the $N=1$ moduli space of vacua
after the perturbation, from the point of view of M theory fivebrane. 
It reproduces the equations
which determine the vevs of massless dyons along the singular locus.
The dyon vevs  $m_{i, dy}^2$, given by (\ref{dyoneven}) $
m_{i,dy}^2 = \sqrt{2} p_i^2 H (p_i^2) \sqrt{T(p_i^2)}, $
are nothing but the difference between the two finite values of $v^2 w^2$.
This can be seen by taking $v=\pm p_i$ in 
(\ref{nht}).
The $N=2$ curve of (\ref{cur}) and (\ref{curve1}) contains  double points
 at $v=\pm p_i$
and $t=C(p_i^2)$. The perturbation $\Delta W$ of  (\ref{superpo}) splits
 these into
 separate points in $(v, t, w)$ space and the difference in $v^2w^2$ between
 these points  becomes the dyon vevs.
 This is a geometric interpretation of dyon vevs
in M theory brane configuration.
By noting that $w^{2}$ satisfies $ 
w^4 -2 N w^2 + N^2 - T H^2 = 0,$
and restricting the form of $N, T$ and $H$ like as $N \sim c_{1}v^2 + c_{2}, 
T \sim c_{3} v^8 + c_{4} v^{6} + c_{5} v^{4} + c_{6} v^2 +c_7,
H \sim \frac{c_{8}}{v^{2}}$, it leads to $
w^4 +  (c_9 + c_{10} v^2) w^2 +c_{11} =0 $
for some constants $c_{i} ( i = 1,\cdots, 11 )$.
Then we can solve for $v^2$ in terms of $w^2$ to reproduce the result of
\cite{aotaug}.
As all the couplings $\mu_{2k}$ are becoming very large, $H(v^2)$ and $N(v^2)$
go to infinity. The term of $N^2-T H^2$ goes to zero as we take the limit
of $\Lambda_{N=2} \rightarrow 0$. This tells us that $w^2$ becomes
$( N^2-T H^2 )/2N$ and as $N(v^2)$ goes to zero, $w^2 \rightarrow \infty$
showing the findings in \cite{egk}.

\section{The Meson Vevs in M Theory  }
\setcounter{equation}{0}

We continue to study for the meson vevs from the singularity structure
of $N=2$ Riemann surface. The vevs of meson will depend on the moduli structure
of $N=2$ Coulomb branch ( See, for example, (\ref{wi2}) ).
Also, the finite values of $w^2$ can be determined fully by using the property of
boundary conditions of $w^2$ when $v$ goes to be very large.
Let us consider the case of finite $w^2$ at $t=0, v=\pm m_i$ and we want 
to compare with the meson
vevs we have studied in (\ref{meson}).
At a point where there exists a single massless dyon ( in other words,
by putting $l=1$ )
and recalling the definition of $T(v^2)$, we have for Yang-Mills with  matter
\bea 
 \left( v^2 C_{2N_c}(v^2)+\Lambda_{N=2}^{2N_c+2-N_f} \prod_{i=1}^{N_f} m_i \right)^2-
\Lambda_{N=2}^{4N_c+4-2N_f}  \prod_{i=1}^{N_f}(v^2-m_i^2) = 
(v^2-p_1^2)^2 T(v^2),
\label{onedyon}
\eea
and the function $w^2$ according to (\ref{nht})  reads
\bea
w^2 = \left[ \frac{h}{v^2} \sqrt{T(v^2)} \right]_+ \pm  \frac{h}{v^2} \sqrt{T(v^2)},
\eea
where in this case $l=1$ means that the polynomial $v^2 H(v^2)$ has the degree of zero and
we denote it by a constant $h$\footnote{ From (\ref{onedyon}) we see
for $N_f < 2N_c+2, 
\sqrt{T(v^{2})}/v^{2} = C(v^{2})/v^{2}(v^{2}-p_1^{2}) + {\cal O}(v^{-4}) $
and we decompose $C$ as $
C(v^{2})/v^{2} = C(p_1^{2})/p_1^{2} + (v^2-p_1^2)\tilde{C}(v^2)$
for some polynomial  $v^2 \tilde{C}(v^2)$ of degree $2N_c$.
This means that the coefficients of $\tilde{C}(v^2)$ can be fixed from 
the explicit form of the polynomial
$C(v^2)$.
The part with nonnegative powers 
of $v^2$ in
$\sqrt{T(v^{2})}/v^{2}$ becomes $\tilde{C}(v^2)$
as follows $
\sqrt{T(v^{2})}/v^{2} = \tilde{C}(v^2) + 
{\cal O}(v^{-2}) \longrightarrow \left[\sqrt{T(v^{2})}/v^{2} \right]_+ = 
\tilde{C}(v^2) $.}.
Thus as $v \rightarrow \pm m_i$ the finite value of $w^2$, 
denoted by $w_i^2$ can be written as
\bea 
w^2_i = w^2(v^2 \rightarrow m_i^2) = 
h \tilde{C} (m_i^2)  \pm  \frac{h}{m_i^2} \sqrt{T(m_i^2)}.
\label{wi}
\eea From (\ref{onedyon}), the relation
$\sqrt{T(m_i^2)}/m_i^2 =  C(m_i^2)/m_i^2(m_i^2-p_1^2)+ 
{\cal O}(m_i^{-4})$
holds 
and the decomposition of  $C$ yields  $
\sqrt{T(m_i^2)}/m_i^2 =  C(p_1^2)/p_1^2(m_i^2-p_1^2) + 
\tilde{C}(m_i^2). $
By plugging this value into (\ref{wi}) and taking the minus sign 
which corresponds to
$t \rightarrow 0$,
we end up with
\bea 
w_i^2 =  \frac{h}{p_1^2} \frac{C(p_1^2)}{(p_1^2-m_i^2)} =
 \frac{h}{p_1^2} \Lambda_{N=2}^{ 2N_c+2 - N_f} \frac{\det 
( p_1^2 - m^2)^{1/2}}{(p_1^2-m_i^2)},
\label{wi2}
\eea
where  we evaluated $ C(p_1^2)$  from (\ref{onedyon})
at $v^2=p_1^2$.
In the above expression we need to know the values of $h$ and $p_1$.
The boundary condition for $w^2$ for large $v$ leads to
\bea
w^2 \sim 2 h \frac{ C_{2N_c}(v^2)}{v^2-p_1^2} 
\sim 2 h v^{2(N_c-1)} + 2 h p_1^2 v^{2(N_c-2)} + 
\cdots,
\eea
which should be equal to
$ \sum_{k=1}^{N_c} 2k \mu_{2k} v^{2(k-1)}$.
Now we can read off the values of $h$ and $p_1$ by comparing both sides,
\bea
h =  N_c \mu_{2N_c} , 
\qquad p_1^2 =\frac{(N_c-1) \mu_{2(N_c-1)}}{N_c \mu_{2N_c}}.
\eea
Finally, the finite value for $w^2$ can be written as
\bea
w_i^2 =\frac{N_c^2 \mu_{2N_c}^2}{(N_c-1) \mu_{2(N_c-1)} }
\Lambda_{N=2}^{2N_c+2-N_f}  
\frac{\det ( a_1^2 - m^2)^{1/2}}{(a_1^2-m_i^2)},
\eea
which is exactly, up to constant, the same expression for 
meson vevs (\ref{meson}) obtained from field theory analysis
in the low energy superpotential (\ref{lowsuperpo}). This illustrates the fact that
at vacua with enhanced gauge group $SU(2)$ the effective superpotential by integrating
in method with the assumption of $W_{\Delta}=0$ is really exact.

\section{Discussions }
\setcounter{equation}{0}

It is straightforward to deal with Yang-Mills theory with massless matter.
When some of branch points of (\ref{curve}) collide as one changes the moduli,
the Riemann surface degenerates and gives a singularity in the theory corresponding
to an additional massless field. By redefining $y$ we get the $2r+1$ branch points of
Riemann surface and one may expect that an unbroken $Sp(r)$ gauge symmetry.

It is easy to generalize the case of several massless dyons. The classical moduli
space is given by several eigenvalues of $a_i$'s. After integrating out the adjoint fields
in each $SU(r_i)$, we obtain the scale matching condition between the high energy scale
and the low energy scale. Then the meson vevs can be written as the differentiation
of the low energy effective superpotential with respect to $m_i^2$ in field
theory approach. On the other hand, in the M theory fivebrane
configuration we can proceed the method done in a single massless dyon.
That is, $\sqrt{T(v^2)}/v^2$ should be expressed for the several dyons and
the decomposition of $C$ will be given also. 
One realizes that a mismatch is found between
field theory results and brane configuration results with $W_{\Delta}=0$. This tells us
that the minimal form for the effective superpotential obtained by integrating in method
is not exact. That is, $W_{\Delta}$ is not zero for the enhanced gauge group $SU(r), r > 2$.
It will be interesting to find the corrections in the future.

\end{document}